\documentclass[a4paper]{jpconf}
\usepackage{graphicx}
\begin{document}
\title{Relativistic U(3) Symmetry and Pseudo-U(3) Symmetry of the Dirac Hamiltonian}

\author{J. N. Ginocchio}

\address{Theoretical Division, Los Alamos National Laboratory, Los Alamos, NM, 87545, USA}

\ead{gino@lanl.gov}                                                                                                                                                                                                                                                                                                                                                                                                                                                                                                                                                                                                                                                                                                                                                                                                                                                                                                                                                                                                                                                                                                                                                                                                                                                                                                                                                                                                                                                                                                                                                                                                                                                                                                                                                                                                                                                                                                                                                                                                                                                                                                                                                                                                                                                                                                                                                                                                                                                                                                                                                                                                                                                                                                                                                                                                                                                                                                                                                                                                                                                                                                                                                                                                                                                                                                                                                                                                                                                                                                                                                                                                                                                                                                                                                                                                                                                                                                                                                                                                                                                                                                                                                                                                                                                                                                                                                                                                                                                                                                                                                                                                                                                                                                                                                                                                                                                                                                                                                                                                                                                                                                                                   
                                                       
\begin{abstract}                      
The Dirac Hamiltonian with relativistic scalar and vector harmonic oscillator potentials has been solved analytically in two limits. One is the spin limit for which spin is an invariant symmetry of the the Dirac Hamiltonian and the other is the pseudo-spin limit for which pseudo-spin is an invariant symmetry of the the Dirac Hamiltonian. The spin limit occurs when the scalar potential is equal to the vector potential plus a constant, and the pseudospin limit occurs when the scalar potential is equal in magnitude but opposite in sign to the vector potential plus a constant. Like the non-relativistic harmonic oscillator, each of these limits has a higher symmetry. For example, for the spherically symmetric oscillator, these limits have a U(3) and pseudo-U(3) symmetry respectively. We shall discuss the eigenfunctions and eigenvalues of these two limits and derive the relativistic generators for the  U(3) and pseudo-U(3) symmetry. We also argue, that, if an anti-nucleon can be bound in a nucleus, the spectrum will have approximate spin and U(3) symmetry.
\\\end{abstract}

\section{Introduction}

Pseudopsin symmetry is approximately conserved in nuclei \cite {gino05}. Pseudospin doublets have single nucleon quantum numbers $[n,\ell,\ell+ {1\over 2}]$ and $[n-1,\ell+2, \ell+{3\over 2}]$ where $n,\ell$ are the radial and orbital quantum numbers, respectively, and the last quantum number in the brackets is the total angular momentum, $j$. For example, in the usual notation, $(1s_ {1\over 2},0d_ {3\over 2}), (1p_ {3\over 2},0f_ {5\over 2})$, etc, are pseudospin doublets. These doublets are approximately degenerate in energy. Some beautiful examples have been shown in this conference; for example, in the talk by O. Sorlin \cite {sorlin08}. Most of the properties of nuclei are explained by the non-relativistic shell model \cite {talmi}. However, to explain pseudospin symmetry we need the relativistic Dirac Hamiltonian which has two types of mean fields, a Lorentz scalar $V_S(\vec r)$ and Lorentz vector $V_V(\vec r)$. Pseudospin symmetry occurs when the sum of the vector and scalar potentials in the Dirac Hamiltonian is a constant , $V_S(\vec r)$ + $V_V(\vec r)= C_{ps}$ \cite {gino97}. Hence pseudospin symmetry is a relativistic symmetry. The approximate equality in magnitude of the vector and scalar fields in nuclei and their opposite sign have been confirmed in relativisitc mean field theories \cite {gino05} and in QCD sum rules \cite {cohen}.

On the other hand hadrons \cite {gino05} and anti-nucleons in a nuclear environment have spin symmetry \cite {gino99}. These are relativistic systems and normally, in such systems, we would expect large spin-orbit splittings. However spin symmetry occurs when the difference of the vector and scalar potentials in the Dirac Hamiltonian is a constant \cite {bell75}, $V_S(\vec r)$ - $V_V(\vec r)= C_{s}$, even though these systems are very relativistic.

The non relativistic harmonic oscillator has been useful in the shell model of nuclei. However, the relativistic mean field has been important for  pseudospin in nuclei and for mesons with one heavy quark \cite{page01} and for an anti-nucleon in a nuclear environment \cite{bur02}. For these reasons the relativistic harmonic oscillator has been solved in the tri-axial, axially deformed and spherical limit \cite {gino04}. Like the non-relativistic harmonic oscillator, each of these limits has a higher symmetry. For example, for the spherically symmetric oscillator, these limits have a U(3) and pseudo-U(3) symmetry respectively. We shall discuss the eigenfunctions and eigenvalues of  these two limits and derive the relativistic generators for the  U(3) and pseudo-U(3) symmetry.

\section{The Spherically Symmetric Dirac Hamiltonian with Spin Symmetry}
The Dirac Hamiltonian for a spherical harmonic oscillator with spin symmetry is 

\begin{equation}
H ={\vec \alpha}\cdot {\vec p}
+ \beta M + (1+\beta)V( r),
\label {dirac}
\end{equation}
where ${\vec \alpha}$, $\beta$ are the Dirac matrices,  ${\vec p}$ is the momentum, $M$ is the mass, ${\vec r}$ is the radial coordinate, $r$ its magnitude, and the velocity of light is set equal to unity, c =1.
The generators for the spin SU(2) algebra and the orbital angular momentum SU(2) algebra,
${{\vec S}},{\vec L}$,
which commute with the
Dirac Hamiltonian with any potential $V(r)$, $[\,H\,,\, {\vec S}\,] = [\,H\,,\,{\vec L}\,] =0$, 
are given by \cite {ami}
\begin{equation}
{\vec S} =
\left (
\begin{array}{cc}
 {\vec s}   & 0 \\
0 & U_p\ {\vec  s}\ U_p
\end{array}
\right ), \ 
{\vec L} = \left (
\begin{array}{cc}
{\vec { \ell}}   & 0 \\
0 & U_p\  {\vec { \ell}}\  U_p
\end{array}
\right ),
\label{gen}
\end{equation}
where
$ {\vec s} = {\vec \sigma}/2$ are the usual spin generators,
${\vec \sigma}$ the Pauli matrices, $ {\vec { \ell}} = {({\vec r }\times {\vec p})\over \hbar}$, and
$U_p = \, {\mbox{\boldmath ${\vec \sigma}\cdot {\vec p}$} \over p}$ is the
helicity unitary operator introduced in \cite {blo95}.
\subsection{The Dirac Hamiltonian with Spin Symmetry and Harmonic Oscillator Potential}

With the harmonic oscillator potential $V(r) ={{M\omega^2}\over 2}\ r^2$ the eigenvalue equation is  \cite{gino04}
\begin{equation}
\sqrt{{ E}_{N} + M}\ ({E}_{N} - M) = \sqrt{ 2\ \hbar^2 \omega^2 M}\ (N + {3\over2})
\label {E}
\end{equation}
where $N= 2n + \ell$, is the total harmonic oscillator quantum number, $n$ is the radial quantum number and $\ell$ is the orbital angular momentum. Hence the eigenenergies have the same degeneracies as the non-relativistic harmonic oscillator. This
eigenvalue equation is solved with Mathematica,
\begin{equation}
E_{N} =  { M \over 3} \left [3 B(A_{N}) + {1 } + {4\over 3\ B(A_{n_1,n_2,n_3})} \right]  ,
\label {E1}
\end{equation}
where $B(A_{N}) =   \left [{ A_{N} + \sqrt{A_{N}^2 - {32 \over 27}}\over 2} \right]^{{2\over 3}}$, and $A_{N} = {\sqrt{2}\hbar \omega\over M} (N+ {3 \over 2})$. The spectrum is non-linear in contrast to the non-relativistic harmonic
oscillator; i.e., the relativistic harmonic oscillator is not harmonic. However for small $A_{N}$ (large $M$),
the binding energy, 
\begin{equation}
E_{N} - { M} \approx { M}\ (  {A_{N} \over \sqrt{2}} + \cdots) \approx \hbar \omega \ (N+ {3 \over 2}),
\label {pert}
\end{equation}
in agreement with the
non-relativistic harmonic oscillator. For large $A_{N}$ (small $M$) the spectrum goes as
\begin{equation}
E_{N} \approx { M}\ (  A_{N}^{2\over 3} + {1 \over 3} + \cdots) ,
\label {asymp}
\end{equation}
which, in lowest order, agrees with the spectrum for ${M}\rightarrow$  0 \cite{bahdri}.

\subsection{U(3) Generators}

The relativistic energy spectrum has the same degeneracies as the non-relativistic spectrum, even though the dependence on $N$ is different. This suggests that the relativistic harmonic oscillator has a higher U(3) symmetry.
The non-relativistic U(3) generators are  the orbital angular momentum $ {\vec { \ell}}$, the quadrupole operator $q_m = {1\over  \hbar M\omega}{\sqrt{3\over 2} } (2M^2 {\omega}^2[rr]_m^{(2)}  + {[pp]_m^{(2)}}$),
where $[rr]_m^{(2)}$ means coupled to angular momentum rank 2 and projection $m$, and the total oscillator quantum number operator, ${\cal N}_{NR} = {1\over 2\sqrt{ 2}  \hbar M\omega  } (2M^2 {\omega}^2r^2  + p^2)- {3\over 2}$. They form the closed U(3) algebra 
\begin{equation}
[{ \cal N}_{NR},{\vec { \ell}}] = [ {\cal N}_{NR},q_m] = 0, 
\end{equation}
\begin{equation}
[ {\vec { \ell}},{\vec { \ell}}]^{(t)} = - \sqrt{2}\  {\vec { \ell}}\ {\delta}_{t,1} ,\  [ {\vec { \ell}},q]^{(t)} = -\sqrt{6}\  {q_m}\ {\delta}_{t,2}, \ [q,q]^{(t)} =  3\sqrt{10}\ {\vec { \ell}}\ \delta_{t,1},
\label {cr}
\end{equation}
with $ {\cal N}_{NR}$ generating a U(1) algebra whose eigenvalues are the total number of quanta $N$ and ${\vec { \ell}}, q_m$ generating an SU(3) algebra.
In the above we use the coupled commutation relation between two tenors, $T_1^ {(t_1)},T_2^ {(t_2)}$ of rank $t_1,t_2$ which is  $[T_1^ {(t_1)},T_2^ {(t_2)}]^{(t)}=  [T_1^ {(t_1)}T_2^ {(t_2)}]^{(t)}- (-1)^{t_1+t_2-t}[T_2^ {(t_2)}T_1^ {(t_1)}]^{(t)}$ \cite{french}.

The relativistic orbital angular momentum generators ${\vec L}$ are given in Eq. (\ref{gen}).  We shall now determine the the quadrupole operator $ Q_m$ and monopole operator $  {\cal N}$ that commute with the Hamiltonian in Eq.(\ref {dirac}).  
In order for the quadrupole generator 
\begin{equation}
{ Q_m} =
\left (
\begin{array}{cc}
 {(Q_m)_{11}}   &  {(Q_m)_{12}} {\vec \sigma}\cdot {\vec p}\\
 {\vec \sigma}\cdot {\vec p}\ {(Q_m)_{21}}  &  {\vec \sigma}\cdot {\vec p}\  {(Q_m)_{22}}\   {\vec \sigma}\cdot {\vec p}
\end{array}
\right ), 
\label{qgen}
\end{equation}
to commute with the Hamiltonian, $[Q_m,H] = 0$, the matrix elements must satisfy the 
conditions,

 \begin{equation}
 {(Q_m)_{12}} = {(Q_m)_{21}},
 \end{equation}
 \begin{equation}
2[{( Q_m)_{11}} ,V]+[{(Q_m)_{12}},p^2] = 0,
 \end{equation}
\begin{equation}
2[{( Q_m)_{12}} ,V]+[{(Q_m)_{22}},p^2] = 0,
 \end{equation}
 \begin{equation}
 {(Q_m)_{11}}= {(Q_m)_{12}}\ 2(V+M) + {(Q_m)_{22}}\  p^2.
  \label{qme}
 \end{equation}

One solution is
\begin{eqnarray}
{ Q_m} =
\lambda_2\ \left (
\begin{array}{cc}
 {{M\omega^2}}\  ({{M\omega^2}}\ r^2+2M) [rr]^{(2)}_m\ + [pp]^{(2)}_m&\ \ \  {{M\omega^2}\ [rr]^{(2)}_m}\   {\vec \sigma}\cdot {\vec p}\\\
{\vec \sigma}\cdot {\vec p}\ {{M\omega^2}}\ [rr]^{(2)}_m  & \ \ \ [pp]^{(2)}_m
\end{array}
\right ), 
\label{Q}
\end{eqnarray}
where $\lambda_2$ is an overall constant undetermined by the commutation of $Q_m$ with the Dirac Hamiltonian. 

For this quadrupole operator to form a closed algebra, the commutation with itself must be the orbital angular momentum operator as in Eq. (\ref{cr}). This commutation relation gives
\begin{equation}
[Q,Q]^{(t)} =\sqrt{10}\  \lambda_2^2\ {{M\omega^2}}{\hbar}^2\ \left (
\begin{array}{cc}
 ({{M\omega^2}}\ r^2+2M) \ {\vec \ell} & \  \ {\vec \ell}\  {\vec \sigma}\cdot {\vec p}\\
{\vec \sigma}\cdot {\vec p}\ {\vec \ell} & \  \ 0
\end{array}\right )= \sqrt{10}\  \lambda_2^2\ {{M\omega^2}}{\hbar}^2\\ (H +M)\ {\vec L}\  \delta_{t,1},
 \label {crq}
\end{equation}
and we get the desired result if $\lambda_2 =  \sqrt{ 3 \over {{M\omega^2}} {\hbar}^2(H+M)}$.
The quadrupole operator then becomes

\begin{eqnarray}
{ Q_m} =
  \sqrt{ 3 \over {{M\omega^2}} {\hbar}^2(H+M)}\ \left (
\begin{array}{cc}
 {{M\omega^2}}\  ({M\omega^2}\ r^2+2M) [rr]^{(2)}_m\ + [pp]^{(2)}_m&\ \ \ {{M\omega^2}}\ [rr]^{(2)}_m\ {\vec \sigma}\cdot {\vec p}\\\
{\vec \sigma}\cdot {\vec p}\ {{M\omega^2}}\ [rr]^{(2)}_m  &\ \ \  [pp]^{(2)}_m
\end{array}
\right )
\label{Q}
\end{eqnarray}
which can also be written as
\begin{equation}
{ Q_m} = \sqrt{ 3 \over {{M\omega^2}} {\hbar}^2(H+M)}\ \left (M \omega^2\ (H + M)\ [rr]^{(2)}_m + [pp]^{(2)}_m \right ),
\label{QS}
\end{equation}
which reduces to the non-relativistic quadrupole generator for $H \rightarrow M$. In the original paper that derives the quadrupole generators there are two typos. In Eq (6) of that paper  \cite{ginoprl05}, ${M\omega^2\over 2}\ r^2$ should be replaced by ${M\omega^2}\ r^2$ and in the non-relativistic quadrupole operator $M^2 {\omega}^2[rr]_m^{(2)} $ should be replaced by  $2M^2 {\omega}^2[rr]_m^{(2)} $. Also, the expression for $B(A_N)$ in that paper has a misplaced factor of 2 in the denominator.

For the monopole  generator, we can solve the same equations.This has been done \cite{ginoprl05}. But there is a simpler way. From Eq (\ref{E}) we get,
\begin{eqnarray}
{ {\cal N}} ={\sqrt{H+M}(H-M)\over \hbar  \sqrt{2M\omega^2}}
- {3\over2}.
\label{NH}
\end{eqnarray}
In the non-relativisitc limit, $H + M\rightarrow 2M$ and the non-relativistic Hamiltonian $(H-M)\rightarrow \hbar \omega (N+ {3\over2})$ which gives the correct result.

The commutation relations are then those of the U(3) algebra, 
\begin{equation}
[{ \cal N},{\vec { L}}] = [ {\cal N},Q_m] = 0. 
\end{equation}
\begin{equation}
[ {\vec { L}},{\vec { L}}]^{(t)} = -\sqrt{2}\  {\vec { L}}\ {\delta}_{t,1} ,\  [ {\vec { L}},Q]^{(t)} =   -\sqrt{6}\  {Q}\ {\delta}_{t,2}, \ [Q,Q]^{(t)} = 3\sqrt{10}\ {\vec { L}}\ \delta_{t,1}.
 \label {crr}
\end{equation}

The spin generators in Eq.(\ref{gen}), $\vec S$, commute with the U(3) generators as well as the Dirac Hamiltonian, and so the invariance group is U(3)$ \times $SU(2), where the SU(2) is generated by the spin generators, $[{\vec S},{\vec S}]^{(t)} = - \sqrt{2}\ {\vec S}\ \delta_{t,1}$.

\section{The Spherically Symmetric Dirac Hamiltonian with Pseudospin Symmetry}

The Dirac Hamiltonian with pseudospin symmetry is \cite {gino97}

\begin{equation}
{\tilde H }={\vec \alpha}\cdot {\vec p}
+ \beta M + (1-\beta)V( r),
\label {pdirac}
\end{equation}
which explains the pseudospin doublets observed in nuclei \cite {gino05}. This pseudospin Hamiltonian can be obtained from the spin Hamiltonian with a transformation  

\begin{equation}
\gamma_5 = \left (
\begin{array}{cc}
0  & 1\\
1 & 0
\end{array}
\right ) and  \   M \rightarrow -M,
\label{tran}
\end{equation}
which gives the pseudospin and pseudo-orbital angular momentum generators  \cite {ami} 
 \begin{equation}
 {\vec {\tilde S} }=
\left (
\begin{array}{cc}
 U_p\ {\vec  s}\ U_p   & 0 \\
0 & {\vec s}
\end{array}
\right ), \ 
{\vec {\tilde L}} = 
\left (
\begin{array}{cc}
U_p\  {\vec { \ell}}\  U_p  & 0 \\
0 & {\vec { \ell}} 
\end{array}
\right ).
\label{pgen}
\end{equation}
\subsection{The Dirac Hamiltonian with Pseudospin Symmetry and Harmonic Oscillator Potential}
With the harmonic oscillator potential $V(r) ={{M\omega^2}\over 2}\ r^2$ the eigenvalue equation in the pseudospin limit is  \cite{gino04}
\begin{equation}
\sqrt{{ E}_{{\tilde N}} - M}\ ({E}_{{\tilde N}} + M) = \sqrt{ 2\ \hbar^2 \omega^2 M}\ ({\tilde N} + {3\over2})
\label {Eps}
\end{equation}
where $ {\tilde N} = 2{\tilde n} + {\tilde \ell}$, is the pseudo total harmonic oscillator quantum number, $ {\tilde n}$ is the pseudo radial quantum number and $ {\tilde \ell}$ is the pseudo-orbital angular momentum. While $ {n}$ is the number of radial nodes  and $ { \ell}$ the rank of the spherical harmonic of the upper Dirac radial amplitude, $ {\tilde n}$ is the number of radial nodes  and $ {\tilde \ell}$ the rank of the spherical harmonic of the lower Dirac radial amplitude. Again the eigenenergies have the same degeneracy pattern as the non-relativistic harmonic oscillator in the spin symmetry limit. This
eigenvalue equation is solved on Mathematica,
\begin{equation}
E_{{\tilde N}} =  { M \over 3} \left [3 B(A_{{\tilde N}}) - {1 } + {4\over 3\ B(A_{{\tilde N}})} \right]  ,
\label {E1}
\end{equation}
where $B(A_{N}) =   \left [{ A_{{\tilde N}} + \sqrt{A_{{\tilde N}}^2 + {32 \over 27}}\over 2} \right]^{{2\over 3}}$, and $A_{{\tilde N}} = {\sqrt{2}\hbar \omega\over M} ({\tilde N}+ {3 \over 2})$. The spectrum is non-linear in contrast to the non-relativistic harmonic
oscillator; i.e., the relativistic harmonic oscillator is not harmonic in either limit. Even for small $A_{{\tilde N}}$ (large $M$),
the binding energy, 
\begin{equation}
E_{{\tilde N}} - { M} \approx { M} (  {A_{{\tilde N}}^2 \over 4} + \cdots),
\label {pert}
\end{equation}
and hence goes quadratically with the the total pseudo-number of quanta and is non-linear even for large $M$.
 For large $A_{{\tilde N}}$ (small $M$) the spectrum goes as
\begin{equation}
E_{{\tilde N}} \approx { M}\ (  A_{{\tilde N}}^{2\over 3} - {1 \over 3} + \cdots) ,
\label {asymp}
\end{equation}
which, in lowest order, agrees with the spectrum for spin symmetry.
\subsection{Pseudo-U(3) Generators}
The pseudo-U(3) generators which commute with the
Dirac Hamiltonian, $[{\tilde H}, {\vec {\tilde S} }] = [{\tilde H}, {\vec {\tilde L} }] =[{\tilde H},{\tilde{Q}_m}] = [{\tilde H},{\tilde{\cal N}}]=0$, 
are then obtained by the transformation in Eq (\ref{tran}) and are given by 
\begin{equation}
{\tilde Q}_m=\sqrt{ 3 \over {{M\omega^2}} {\hbar}^2({\tilde H}-M)}\ \left (M \omega^2\ ({\tilde H}-M)\ [rr]^{(2)}_m + [pp]^{(2)}_m \right )\end{equation}
\label{Hps}
\begin{equation}
{\tilde {\cal N}} =
{\sqrt{{\tilde H}-M}({\tilde H}+M)\over\hbar\sqrt{ 2M\omega^2}}
- {3\over2}
\end{equation}
The commutation relations are then those of the U(3) algebra, 
\begin{equation}
[{\tilde {\cal N}},{\vec {\tilde L}}] = [ {\tilde {\cal N}},{\tilde Q}_m] = 0. 
\end{equation}
\begin{equation}
[ {\vec {{\tilde L}}},{\vec { {\tilde L}}}]^{(t)} = -\sqrt{2}\  {\vec {{\tilde L}}}\ {\delta}_{t,1} ,\  [ {\vec { {\tilde L}}},{\tilde Q}]^{(t)} =   -\sqrt{6}\  {{\tilde Q}}\ {\delta}_{t,2}, \ [{\tilde Q},{\tilde Q}]^{(t)} = 3\sqrt{10}\ {\vec { {\tilde L}}}\ \delta_{t,1}.
 \label {pcrr}
\end{equation}

The pseudospin generators in Eq.(\ref{pgen}), $ {\vec {\tilde S} }$, commute with the U(3) generators as well as the Dirac Hamiltonian, and so the invariance group is pseudo-U(3)$ \times $pseudo-SU(2), where the pseudo-SU(2) is generated by the pseudospin generators, $[{\vec {\tilde S} },{\vec {\tilde S} }]^{(t)} = - \sqrt{2}\ {\vec {\tilde S} }\ \delta_{t,1}$.

\section{The Axially Deformed Dirac Hamiltonian with a Harmonic Oscillator Potential}
The relativistic non-spherical  harmonic oscillator has also been solved analytically in these two limits \cite{gino04}. The  non-relativistic axially symmetric deformed harmonic oscillator will have a U(2)$\times$U(1)  symmetry. Likewise the relativistic axially symmetric deformed  harmonic oscillator will have a U(2)$\times$U(1) symmetry in the spin symmetry limit and a pseudo-U(2)$\times$ pseudo-U(1) symmetry in the pseudospin limit. This will be discussed in a forthcoming paper.  
\section{Pseudospin Symmetry Limit and Perturbation Theory}
Even though the harmonic oscillator eigenenergies for the pseudospin symmetry limit, Eq(\ref{E1}), are  positive, the eigenstates \cite {gino04} are not those of Dirac valence states but of Dirac hole states. This follows from the fact that the eigenstates have the radial nodal structure of hole states rather than valence states. For example, the state with ${\tilde n} = {\tilde \ell} =0$, has by definition a lower amplitude with zero radial nodes. This state corresponds to a $1p_ {{1\over 2}}$ state. Using a theorem which relates the radial nodes of upper and lower amplitudes \cite {gino01},  the lower amplitude of a  $1p_ {{1\over 2}}$ state would have to have one radial node if it were a valence state. Hence this state is a hole state. This means that perturbation theory can not be used to describe nuclei even though nuclei have approximate pseudospin symmetry. In general, the pseudospin limit does not have bound valence states independent of the form of $V(r)$  \cite {gino01} and hence perturbation theory is not possible. For this reason, the analytical solution of the Dirac Hamiltonian with general scalar and vector harmonic oscillator potentials would be useful.
\section{Spin Symmetry Limit and Anti-nucleons}
The fact that a nucleon moving in the mean field of a nucleus has a vector and scalar potential almost equal in magnitude but opposite in sign has been substantiated by experimental evidence in nuclei \cite{gino05}. An anti-nucleon moving in the mean field of a nucleus will then have a vector and scalar potential almost equal in magnitude because charge conjugation leaves the Lorentz scalar field invariant while changing the sign of the Lorentz vector field. Thus pseudospin in nuclei predicts spin symmetry for an anti-nucleon in a nuclear environment  \cite {gino99,gino05}.

\section{Summary}
We have shown that a Dirac Hamiltonian with equal scalar and vector harmonic oscillator potentials has an U(3)$\times$ SU(2) symmetry and with scalar and vector harmonic oscillator potentials equal  in magnitude but opposite in sign has a pseudo-U(3) $\times$ pseudo-SU(2) symmetry and we have derived the corresponding generators for each case. If speculation that an anti-nucleon can be bound inside a nucleus is valid \cite {bur02}, the anti-nucleon spectrum will have an approximate spin symmetry and, most likely, an approximate U(3) symmetry, because the vector and scalar potentials are approximately equal and very large \cite {gino05}.

\ack
This research was supported by the US Department of Energy, Contract No. W-7405-ENG-36.


\section*{References}
\begin{thebibliography}{9}
\bibitem{gino05}
Ginocchio J N 2005  {\it Phys.Rep.} {\bf 414} 165
\bibitem{sorlin08}
Sorlin O and Porquet M G 2008  {\it Prog. Part. Nucl. Phys.} {\bf 61} 602
\bibitem{talmi} 
Talmi I 1993 {\it Simple Models of Complex Nuclei}, (Switzerland: Hardwood )
\bibitem{gino97}
Ginocchio J N 1997 {\it Phys.Rev. Lett.} {\bf 78} 436
\bibitem{cohen}
Cohen T D, {\it et al} 1995 {\it Prog. Part. Nucl. Phys.}  {\bf 35} 221
\bibitem{gino99}
Ginocchio J N 1999  {\it Phys.Rep.} {\bf 315} 231
\bibitem{bell75}
Bell J S and Ruegg H 1997 {\it Nucl.Phys.} B {\bf 98} 151
\bibitem{page01}
Page P R, Goldman T and Ginocchio J N 2001 {\it Phys.Rev. Lett.} {\bf 86} 204
\bibitem{bur02}
Burvenich T, {\it et al} 2002 {\it Phys.Lett.} B {\bf 542} 261
\bibitem{gino04}
Ginocchio J N 2004 {\it Phys.Rev.} C {\bf 69} 034318
\bibitem{ami}
Ginocchio J N and Leviatan A 1998
{\it Phys. Lett.} B  {\bf 425} 1 
\bibitem{blo95}
Blokhin A L, Bahri C and Draayer J P 1995 {\it Phys.Rev. Lett.} {\bf 74} 4149
\bibitem{bahdri}
Bhaduri R K 1988 {\it Models of the Nucleon: From quarks to Soliton}, 
(Addison-Wesley)
\bibitem {french}
French J B 1966
{\it Many-Body Description of Nuclear Structure and Reactions}, ed C. Bloch
(New York: Academic Press). 
\bibitem{ginoprl05}
Ginocchio J N 2005 {\it Phys.Rev.Lett.} {\bf 95} 252501 
\bibitem{gino01}
 Leviatan A and Ginocchio J N 2001 {\it Phys.Lett.} B {\bf 518} 214






\end{thebibliography}
\end{document}